\documentclass[]{spie}

\usepackage[]{graphicx}

\title{Clio: a 3-5 micron AO planet-finding camera} 

\author{Suresh Sivanandam\supit{a}, Phil M. Hinz\supit{a}, Ari N. Heinze\supit{a}, Melanie Freed\supit{b}, Andrew H. Breuninger\supit{a}
\skiplinehalf
\supit{a}Steward Observatory, 933 North Cherry Avenue, Tucson, AZ 85721, USA;\\
\supit{b}College of Optical Sciences, 1630 East University Boulevard, Tucson, AZ 85721, USA
}

\authorinfo{Further author information: (Send correspondence to S.S.)\\
S.S.: E-mail: suresh@as.arizona.edu, Telephone: 1 520 621 1455}

\begin{document} 
\maketitle 

\begin{abstract}
Clio is an adaptive-optics camera mounted on the 6.5 meter MMT optimized for diffraction-limited L' and M-band imaging over a $\sim15''$ field. The instrument was designed from the ground up with a large well-depth, fast readout thermal infrared ($\sim 3-5\:\mu m$) 320 by 256 pixel InSb detector, cooled optics, and associated focal plane and pupil masks (with the option for a coronograph) to minimize the thermal background and maximize throughput. When coupled with the MMT's adaptive secondary AO (two warm reflections) system's low thermal background, this instrument is in a unique position to image nearby warm planets, which are the brightest in the L' and M-band atmospheric windows. We present the current status of this recently commissioned instrument that performed exceptionally during first light. Our instrument sensitivities are impressive and are sky background limited: for an hour of integration, we obtain an L'-band 5 $\sigma$ detection limit of of 17.0 magnitudes (Strehl $\sim 80\%$) and an M-band limit of 14.5 (Strehl $\sim 90\%$). Our M-band sensitivity is lower due to the increase in thermal sky background. These sensitivities translate to finding relatively young planets five times Jupiter mass (M$_{Jup}$) at 10 pc within a few AU of a star. Presently, a large Clio survey of nearby stellar systems is underway including a search for planets around solar-type stars, M dwarfs, and white dwarfs. Even with a null result, we can place strong constraints on planet distribution models.
\end{abstract}


\section{Introduction}

Despite the exhaustive search for extrasolar planets, only one confirmed exoplanet orbiting around the brown dwarf 2M1207 has been imaged through VLT/NACO observations~\cite{2005A&A...438L..25C}. The first detection of exoplanet photons resulted from Spitzer observations of a secondary eclipse in a known planetary system; however, the system was not resolved~\cite{2005ApJ...626..523C}. The contrast levels needed to discern Jupiter-sized planets visibly (or even in the near-IR) from their bright stars typically require next generation hardware. Thus, the $>100$ \emph{indirect} planet discoveries have resulted mostly from high-precision radial velocity (RV) surveys where some have been followed up with transit studies to determine the physical properties of a planetary system~\cite{2005PThPS.158...24M}.  A handful, including Jupiter-sized planet TrES-1~\cite{2004ApJ...613L.153A}, have turned up in transit studies (e.g. TrES, OGLE) and been confirmed. However, RV and transit surveys are inherently biased towards large, close-in planets, leaving lower-mass planets on orbits with wider semi-major axes largely unexplored and their distributions unknown. Direct imaging is well-poised to uncover planets at greater separations from stars and therefore complements current RV techniques. At present, direct imaging can uncover the distribution and properties of extrasolar giant planets (EGP) and place strong constraints on current evolutionary models. The $3-5\:\mu m$ imaging offers a new window for tackling these scientific questions.

\section{3-5 micron adaptive-optics planet imaging}

Much effort has gone into near-IR (J, H, and K-band) planet imaging~\cite{2006astro.ph..1062B} since near-IR sensitivity on the ground is better than at longer wavelengths. Techniques such as simultaneous spectral difference imaging~\cite{2004SPIE.5490..389B} and higher order adaptive optics~\cite{2004SPIE.5490..433O} have been employed to improve contrasts in the near-IR because speckle noise and low Strehls are thorny issues encountered in this wavelength regime. These searches focus mainly on young stellar systems ($\sim 100$ Myr) since the planets are young and hot which translate to favourable near-IR contrasts. However, most nearby stellar systems are relatively old. Theoretical modelling of EGPs indicate non-blackbody spectral energy distributions (SED) (see Figure 1), with a broad emission feature in the 4-5 $\mu m$ range that is persistent as the planet ages~\cite{1997ApJ...491..856B}. A significant portion of the absorbed radiation is re-emitted in this wavelength range where the EGP's atmosphere is relatively transparent~\cite{2004ApJ...609..407B,2003ApJ...588.1121S}. This peak is indeed observed in Juptier's spectrum and in T-type brown dwarfs and is modelled to be universal in all objects with effective temperatures lower than 1200K. In addition, these models show this feature does not depend as strongly on age as does the near-IR flux. However, there has been some contention if this feature is over-predicted due to non-equilibrium effects of atmospheric carbon monoxide~\cite{2004AJ....127.3516G}. During the design phase of our instrument, we have shown through Monte Carlo simulations of a broad age range (100 Myr - 1 Gyr) of planets that the M-band is the most promising ground-based window for planet imaging is in the M-band~\cite{2004SPIE.5492.1561F}. We expect, unlike other imaging searches, to image warm $> 5$ M$_{Jup}$ EGPs about a Gyr old in nearby ($\sim10$ pc) stellar systems. In this paper, we present the measured capabilities of the instrument and show that they typically meet or exceed our original predictions.

\begin{figure}
\begin{center}
\begin{tabular}{cc}
\includegraphics[width=10cm]{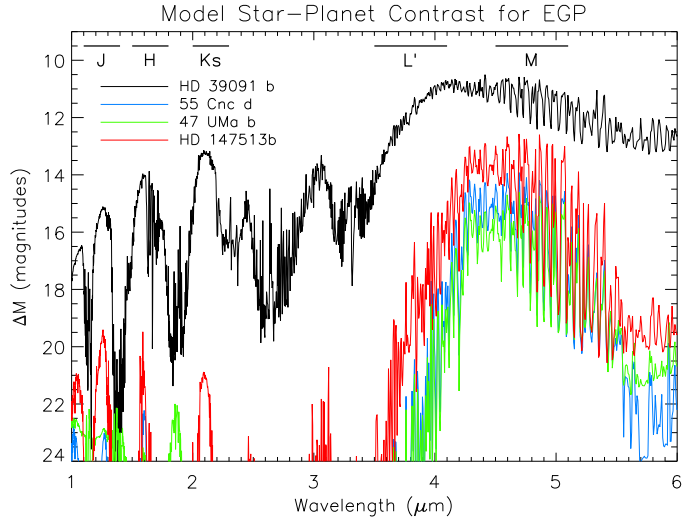}
\end{tabular}
\end{center}
\caption[gainnoise] 
{ \label{fig:gainnoise} Model spectra (in terms of star-planet contrast) of a few known radial-velocity EGPs~\cite{2004ApJ...609..407B,2003ApJ...588.1121S}. The most massive of these, HD 39091b is 10 M$_{Jup}\sin i$ while the remaining cooler planets are in the 1-4 M$_{Jup}\sin i$ range. Host stars are G-type. Bandwidths of near and thermal IR filters are overlaid. Of note is the L' and M-band bump present in every planet spectrum regardless of mass, age, or separation from star. However, the near-IR spectrum is dominant only in young, hot and/or close-in planets where there is significant reflected light.}
\end{figure} 

\section{Instrument description and characterization}

Clio is a 3-5 micron camera optimized to minimize thermal background and maximize throughput in the L' and M-bands while maintaining diffraction-limited performance in these wavebands. The detector has been chosen for high duty-cycle, large well-depth operation due to the high sky background of thermal IR observations. Installed on the 6.5m MMT telescope, this instrument, along with the low emissivity MMT natural guide star (NGS) adaptive optics system, facilitates one-of-a-kind highly sensitive measurements in the thermal IR. Below we discuss the construction of the instrument and laboratory tests that characterize the performance of the detector.

\subsection{Optical Design}

Clio is designed to work at the diffraction limit from H to M-bands and provides Nyquist sampling of the PSF. This broad wavelength range requires two different magnifications to achieve this. Therefore, Clio has 2 lens wheels that work in tandem allowing the camera to operate in f/20 (L' and M) and f/35 (H and K) mode. Since the sensitivity of the detector is poor shortward of 3 $\mu m$, the instrument is mainly operated in f/20 mode. In addition, there is a third option to operate in pupil imaging mode which allows the proper alignment of cold stops in the pupil plane. Lenses are made of either AR-coated Cleartran or ZnSe. Figure 2 shows a schematic of the instrument along with a ray trace illustrating its optical functionality. At the entrance window is a calcium fluoride dichroic that reflects light shortward of 1 $\mu m$ into the visible light wavefront sensor exterior to the instrument. At the secondary focus there are a number of field stops that can be chosen. Coronographic elements can be added here, though there are none at the moment.  The pupil stop allows the selection of several different sized circular stops to match the stop of the telescope preventing any stray thermal emission from entering into the instrument. In addition, PSF shaping phase plates (Codona et al. 2006, this conference) are also included in the pupil stop wheel for further suppression of stellar light.  Following that, a filter wheel assembly consists of two separate wheels that allow the selection of Barr H through M-band filters, including additional filters such as a narrow band M filter, a 3 to 5 $\mu  m$ broadband filter, and an MKO M' filter, the latter of which improves photometric stability. A number of neutral density filters are also included.
\par
Recent observations have verified that Clio does operate at the diffraction limit in the L' and M-bands reaching Strehls, obtained from scaled H-band Strehl, of 0.8 and 0.9 in L' and M-bands respectively for closed-loop stellar AO observations.

\begin{figure}
\begin{center}
\begin{tabular}{cc}
\includegraphics[width=14.5cm]{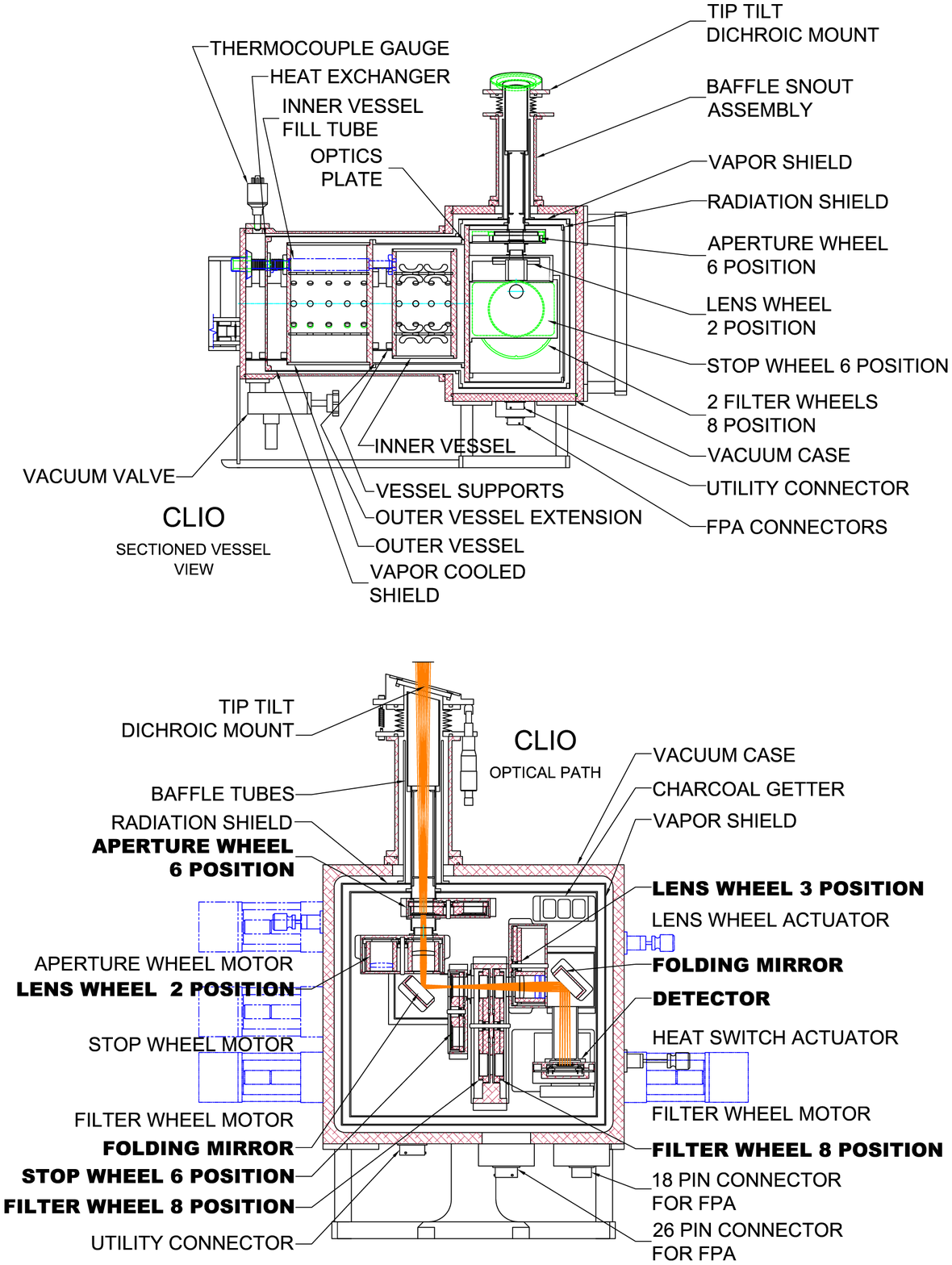}
\end{tabular}
\end{center}
\caption[schematic] 
{ \label{fig:schematic} \textbf{Top:} Profile of Clio cryostat (3/4 size of bottom view). \textbf{Bottom:} A view of the Clio optics within the cryostat. Instrument is shown in f/20 mode; important optical components highlighted. The secondary focus of the telescope is located at the aperture stop wheel which can contain coronographic masks. The first and second lens wheels allow for f/20, f/35, and pupil imaging modes. Pupil stop wheel contains pupil masks to match the entrance pupil of the camera with the telescope pupil. Additional optics such as wave plates and coronographic masks can be installed here. The filter wheel assembly consists of 2 separate wheels with a collection of IR and neutral density filters. }
\end{figure}

\subsection{Cryogenic Dewar}

The cryogenic dewar (Figure 2) was manufactured by IR Labs and was designed from the ground up to minimize instrument emissivity. The dewar consists of 2 cryogen reservoirs, one which cools the optics, and the other cools the detector. In addition, the dewar consists of two radiation shields that keep it light-tight from room-temperature radiation. Additional baffling is added to all of the optics to minimize any stray emission that is not properly shielded. All of the optics are kept at liquid nitrogen (LN2) temperature (77K). The detector's LN2 reservoir is attached to a vacuum pump that reduces the LN2 vapour pressure to a point where it solidifies. We have been able to cool the detector down to 56K using this technique, thereby reducing its dark current by a few orders of magnitude. The pump remains operational throughout the observation to maintain sub-LN2 temperatures. A heater is attached to the detector assembly, which coupled with a thermal sensor and a PID temperature controller, maintains a stable operating temperature for the detector. When operating cold with the entrance window blanked off, we find that the detector signal is dark current dominated indicating the optics makes a negligible contribution to the background.

\subsection{Detector and Control Electronics}

The detector was purchased from Indigo Systems Inc. The detector consists of a $320\times256$ pixel Indium Antimonide (InSb) array bump-bonded onto the ISO9809 readout (ROIC). This particular combination was chosen for its high well-depth (detailed below) - well-depth is critical since typical detectors saturate very quickly in the M-band due to the high sky flux - its ability for rapid readout, and relatively low dark current. The mode of operation is to take short exposures very quickly and coadd them to obtain a high signal-to-noise image despite the high sky background.
\par
This ROIC has a number of features that make it suitable for our application. It operates with 4 readouts where every fourth column of the array is multiplexed into one readout amplifier. With this ROIC we are able to sample-and-hold pixel values, carry out a global reset, and readout the array while it continues integrating. This greatly improves the usability of the array at short integration times since the duty cycle is still fairly high. In addition, at the default low gain setting, the ROIC offers a well-depth of approximately 3 million $e^-.$ The ROIC has a number of power modes, which allow the control of dark current. Its main disadvantage is its high readnoise of 350 $e^-$ (77K) at the low gain setting. However, for our high background application the read noise can be considered negligible. 
\par
The detector is run by SDSU Gen II electronics with a modified timing board and DSP code that allows for faster clocking of the array, though an unmodified timing board should still be adequate. The pixel clock is set at 200 kHz, and pixels are transmitted at each transition edge giving a pixel rate of 400 kHz. The final version of the DSP code implements rapid readouts of the array with integration times as short as 65 ms, though different versions of the DSP code allow even shorter integration times at the expense of duty cycle. For typical M-band integration times of 100 ms, the set-up operates at a 92\% duty cycle, and this improves for longer L'-band integrations. 
\par
Typical operation of the detector is at 56K. Even though the detector is rated for LN2 temperature, we find it is stable at solid nitrogen temperatures with the exception of increased column noise. The salient properties of the detector largely depend on the detector bias voltage. We chose a voltage that maximized well-depth, and minimized dark current and bad pixels. Table 1 describes all of the properties of the detector derived from lab tests. Quantum efficiency measurements were done at the manufacturer's lab over a 3-5 $\mu m$ range. Gain was measured by exposing the array to a low-level of light and by determining the slope of the DN versus $\sigma^2_{DN}$ plot (Figure 3 - Left Panel), and the range of gain values for the 4 readouts is reported here. Gain values were fit over the 15-90\% full-well range. There is a 5\% RMS variation in the gain values of the readouts. Fit slopes had an error of 1\%, which we took to be the linearity of the detector. 200 low light level 200 ms images taken at 3 Hz were used to analyse time variability of individual pixels. Figure 3 (Right Panel) reveals low-level column noise and a few ($\sim0.5\%$) highly variable pixels. Dark current measurements were taken by placing a cold cap, held at the temperature of the detector, on the detector assembly. We believe the majority of dark current is generated by glowing electronics in the detector. We set the detector to operate at the lowest power mode, which decreased dark current an order of magnitude to a point where the array saturated in approximately 10 seconds. We computed the read noise by extrapolating $\sigma^2_{DN}$ for a zero second exposure. We found the read noise to be a factor of 2 higher than the manufacturer's quoted value.

\begin{figure}
\begin{center}
\begin{tabular}{cc}
\includegraphics[width=8cm]{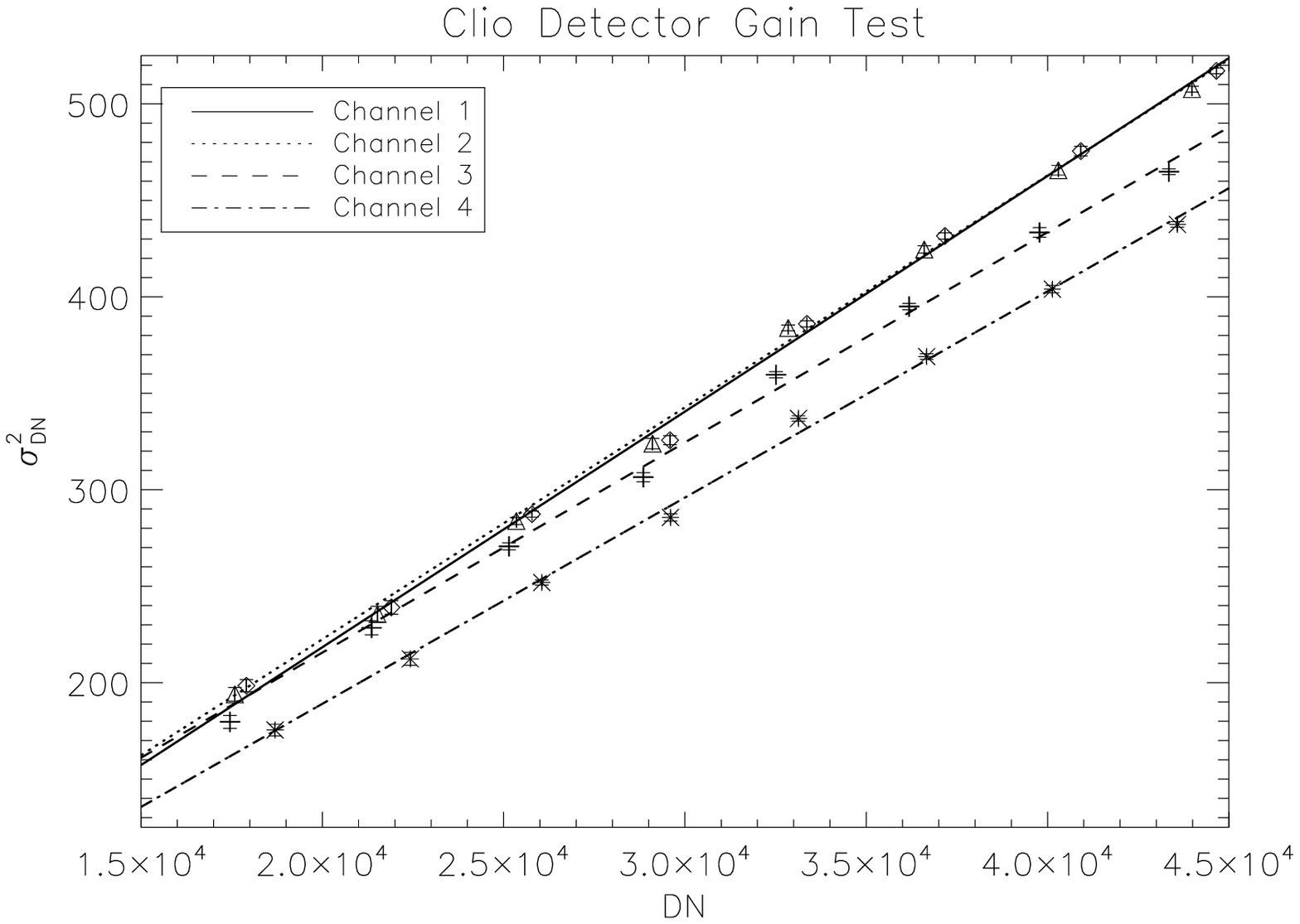} & \includegraphics[width=8cm]{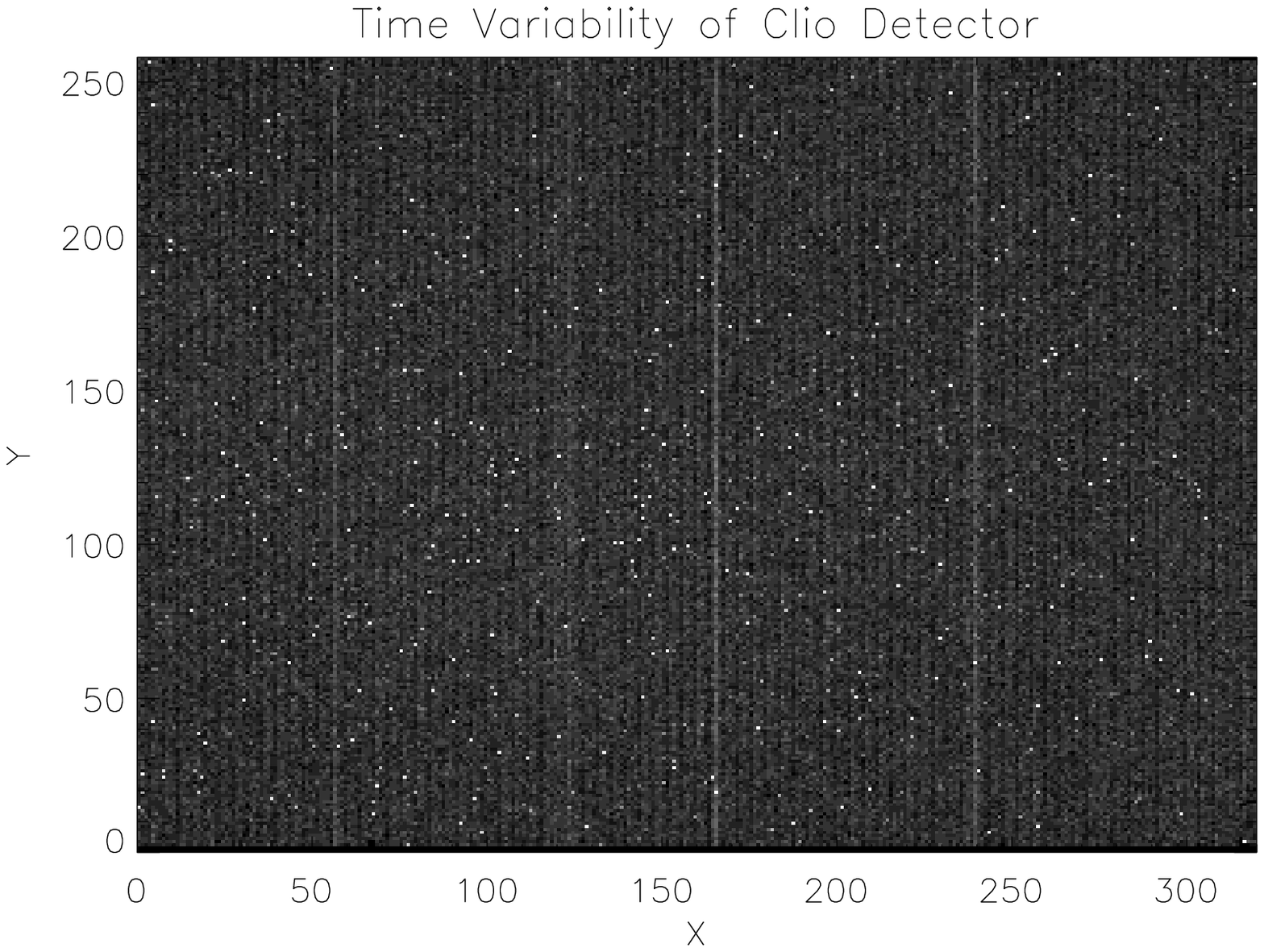}
\end{tabular}
\end{center}
\caption[gainnoise] 
{ \label{fig:gainnoise} \textbf{Left:} Gain fit to all four readout channels. The absicca are raw DN values since we only care about the slopes. All channels are linear to about one percent. Error bars indicate variations in noise over the set of exposures used for each data point. \textbf{Right:} Detector noise map. The image stretch is from 10 to 25 DN. This figure shows the standard deviation per pixel for a 200 ms low light level exposure. We see a random distribution of highly variable pixels along with some low level column noise discussed earlier. All of these artifacts are removed in the image processing pipeline.}
\end{figure} 

\begin{table}[t]
\begin{center}
\caption{Clio Detector Parameters}
\vspace{3mm}
\begin{tabular}{|c|c|}
\hline
\bf{Array size} & $320\times256$ \\
\hline
\bf{Pixel size} & $30\:\mu m$ square \\
\hline
\bf{Bias voltage} & 4.0 V \\
\hline
\bf{Pixel rate} & 400 kHz \\
\hline
\bf{Operating temperature} & 56K \\
\hline
\bf{QE$_{3-5 \mu m}$} & 0.9 \\
\hline
\bf{Full well depth} & $3.3\times10^6$ $e^-$ \\
\hline
\bf{Dark current} & $3.0\times10^5$ $e^-/s$ \\
\hline
\bf{Read noise} & 800 $e^-$ \\
\hline
\bf{Gain} & 82 - 94 $e^-/DN$ \\
\hline
\bf{Linearity} & 1\% \\
\hline

\end{tabular}
\end{center}
\end{table}

\subsection{MMT Adaptive Optics}
The MMT adaptive optics system (MMTAO) is the only design that uses a deformable f/15 secondary that is fully integrated into the telescope~\cite{2003SPIE.5169...17W,2003SPIE.5169...26B}. As a consequence, MMTAO has only two warm reflections (the primary and secondary) while other AO systems have several warm mirrors in their optical paths. Our approach minimizes the emissivity of the telescope structure, essential for better sensitivity in the thermal IR. In addition, the secondary mirror is slightly undersized and serves as the stop for the system resulting in an effective telescope diameter of 6.35 m. This has been shown to reduce the telescope thermal background further. The Shack-Hartman NGS wavefront sensor operates off the reflected visible light from the dichroic at 550 Hz and corrects for up to 56 Zernike modes of atmospheric aberration. A study has shown that this set-up along with an instrument optimized to minimize thermal background, for typical values at the MMT, can reach a sensitivity limit in one-third or one-half of the time of conventional AO systems~\cite{2000PASP..112..264L}.

\section{On-sky Tests}
We carried out several on-sky tests to determine the astronomical performance of the instrument. This involved determining the optimal mode of observations, observing a number of standard stars to characterize sensitivity and throughput, and observing double stars to determine plate scale. All observations were carried out in f/20 mode with the instrument derotator fixed. In pupil imaging mode, we took care to choose and align a cold stop, which matched the stop of the system. Acquired images were processed offline with a pipeline optimized for faint companion detection outlined in Heinze et al.  (2006, this conference). Pertinent details of the instrument derived from these tests are given in Table 2.

\begin{table}[t]
\begin{center}
\caption{Clio Observational Parameters (f/20 mode)}
\vspace{3mm}
\begin{tabular}{|c|c|}
\hline
\bf{Instrument FOV} & $15.5"\times12.4"$ \\
\hline
\bf{Plate scale} & 0.0486"/pix \\
\hline
\bf{Throughput$_{L'}$} & 0.67 \\
\hline
\bf{Throughput$_{M}$} & 0.42 \\
\hline
\bf{L' $5\sigma$ limit (1hr)} & 17.0 mag \\
\hline
\bf{M $5\sigma$ limit (1hr)} & 14.2 mag \\
\hline
\bf{Emissivity} & 10 \% \\
\hline
\end{tabular}
\end{center}
\end{table}

\subsection{Observational Strategy}
The high flux and variable sky background dominates the photon flux in the L' and M-bands and necessitates an effective way of subtracting its flux from the science image. We are effective in removing this background by nodding the telescope by a few arcseconds in either the altitude or azimuth direction after 4 exposures at a particular nod position and subtracting nod pairs. Typical exposure times were 100 ms in M and 1000 ms in L'. The goal was to expose long enough that sky photons filled half the well-depth of the detector. This ensures that all exposures are sky-noise limited. This, however, means that the science object, which is also the guide star for the AO system, could be saturated. Since we are only interested in detecting faint companions and not precise photometry, this is not an issue. In addition, we coadded the images in batches of 20 or more exposures on the acquisition computer as the data were read off the controller. This aided in improving the duty cycle of the readout.
\par
The instrument derotator is fixed and the sky is allowed to rotate with each exposure. This mitigates long-lived speckles that masquerade as faint sources. These speckles are thought to be associated with the static aberrations downstream of the dichroic which are not corrected for by the AO system. We often observe objects that transit through zenith giving us large amounts of rotation. While the sky rotates, we expect the speckles to remain fixed in the image plane since they are associated with static aberrations. We later rotate the images back to the correct orientation during off-line image processing using the object parallactic angle obtained from the AO system. The long-lived speckles are smeared out by this process and no longer appear as possible faint companions. 
\par
During our April 2006 observations, the overall duty-cycle of the above setup including target acquisition and AO loop closure was 70\%.

\subsection{Plate Scale}
We observed two double stars HD 100831 and HD 115404 to determine the plate scale of the instrument. Both observations gave consistent results: $0.048574 \pm 0.000090$ arsec/pixel for HD100831 and $0.048620 \pm 0.000323$ for HD115404. The quoted uncertainties were entirely dominated by the original measurements of the true separation of the binaries and not by the Clio measurement error. The small difference between the derived plate scales suggests that the uncertainties may be overestimated. As a comparison the predicted PSF width for our setup is 0.15" and 0.19" in the L' and M-bands respectively.

Since the instrument derotator is fixed, we need to know the exact orientation of the instrument FOV with respect to the sky at any one time to orient and stack the exposures of a science object. We obtained through these double star observations the image transformation required, which is a function of parallactic angle, to create correct astronomical images with North up and East left. Clio images first need to be mirror flipped right-to-left and rotated $\theta_{rot}$ using the following transformation if the MMT instrument derotator is set to 0 degrees: $\theta_{rot} = 272.98^\circ-\theta_p$ where $\theta_p$ is the parallactic angle for a given exposure. We obtain consistent results for both double stars.

\subsection{Throughput}

We observed an L' and M' standard, HD162208, in order to obtain a photometric calibration and determine the throughput of the system. Leggett et al. quote an L' magnitude of 7.125 and M' magnitude of 7.05 for this star~\cite{2003MNRAS.345..144L}. We observed a L' count rate of $235255 \pm 3775$ DN/s and M count rate of $89273 \pm 2744$ DN/s. Normalizing to a star of 10.0 magnitudes, we obtained $16655 \pm 318$ DN/s for L' and $5898 \pm 181$ DN/s for M. From atmospheric transmission spectrum and stellar SED modelling (9250K blackbody) for this star, we found the colour correction to go from M' to M is negligible. Therefore, we approximated $M \approx M'.$ We computed the throughput for these two observations and obtain a L' value of 0.67 and an M-band value of 0.42. A crude estimate of atmospheric and optics transmission and detector QE is consistent with the derived values. These results indicate our instrument is indeed high-throughput due to the minimum number of reflective and refractive components and a high QE detector.

We observed another standard, HD 162208, on a different night and obtained the following photometric calibrations normalized to a 10th magnitude star: $16126 \pm 214$ DN/s and $4300 \pm 203$ DN/s for L' and M-bands respectively. The L' value is consistent with the previous estimation, though the M' value deviates by 25\%. We believe there was some cloud across the target, making these calibrations unreliable. M-band photometric calibration was likely to be more seriously affected than the L'-band because of the forest of water bands in the M-band window.

\subsection{Sensitivity}

We determined L'-band sensitivity during our most recent observing run by observing GJ 450 and stacking the exposures reaching a total integration time of 5355 s. We obtained a background-limited, 5 $\sigma$ point source detection limit of L' = $17.4 \pm 0.3.$ If we scale this result for a one hour integration, we obtain an L' 5 $\sigma$ source detection limit of $17.0 \pm 0.3.$ Using tabulated theoretical model values~\cite{2003A&A...402..701B}, this corresponds to detecting a 5 M$_{Jup},$ 1 Gyr-old planet and a 10 M$_{Jup},$ 5 Gyr old planet 10 parsecs away. This clearly shows this instrument will be able to set interesting limits on planetary systems of all ages.

We measured the M-band sensitivity during the June 2005 commissioning run by observing Vega. We reached a M-band limiting magnitude of 13.4 for a 672 second exposure~\cite{HinzApJ}. Scaled to a one hour integration, we reach a limiting magnitude of 14.5. This corresponds to 10 M$_{Jup}$, 1 Gyr old planet and a 20 M$_{Jup}$ 5 Gyr old planet 10 parsecs away. However, this was a relatively short integration and may not be representative of the current limiting magnitude because the instrument has since been optimized. Preliminary data reductions from the most recent run suggest the M-band limiting magnitude is approximately 2 magnitudes brighter than that of the L'-band. In light of the L'-M colour of EGPs and improved sensitivity in the L'-band, we find observing in the L'-band provides more sensitive mass constraints and we are focussing our searches in this band. Due to the unique L'-M colour of EGPs, we will carry out M-band follow-up observation to discern the nature of our candidates.

\subsection{Emissivity}

We carried out preliminary emissivity measurements during the April 2006 run. Emissivity measurements were taken in L' and M-band by observing the sky at 1.15 and 2 airmasses, and observing the closed dome. Darks were subtracted from these images to obtain only the incident flux. We assumed the closed dome to act as a black body, and obtained emissivity values, good to about 10\%, by dividing the sky frame flux by the dome frame flux. We find emissivity values of 0.17 and 0.21 in the L'-band at 1.15 and 2 airmasses respectively. We obtain values of 0.31 and 0.37 in the M-band for the same airmasses. If we assume the atmosphere is optically thin, we would expect the sky emissivity to scale linearly with airmass. Using only the L'-band data, we find the telescope+instrument emissivity must be approximately 10\% while the sky emissivity at zenith to be 6\%. This agrees crudely with the atmospheric model predictions of L' transmission. However, the M-band data does not agree with a 10\% emissivity. It may be that the optically thin assumption breaks down in the M-band due to saturation of strong water bands. We caution that this is only a preliminary measurement. We expect our future cold stops will include a central obscuration to further decrease the telescope emissivity.

\begin{figure}
\begin{center}
\begin{tabular}{cc}
\includegraphics[width=8.25cm]{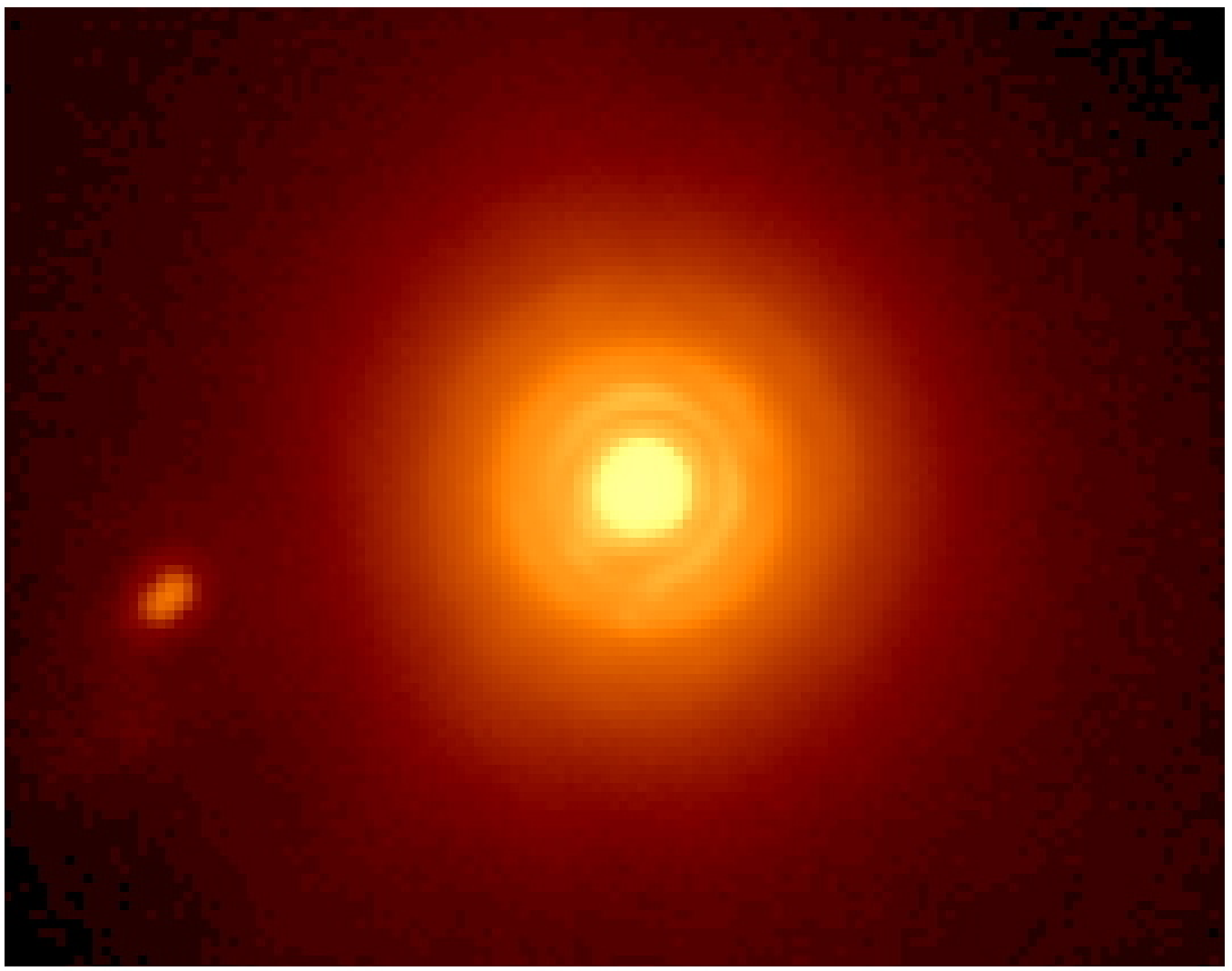} & \includegraphics[width=7.75cm]{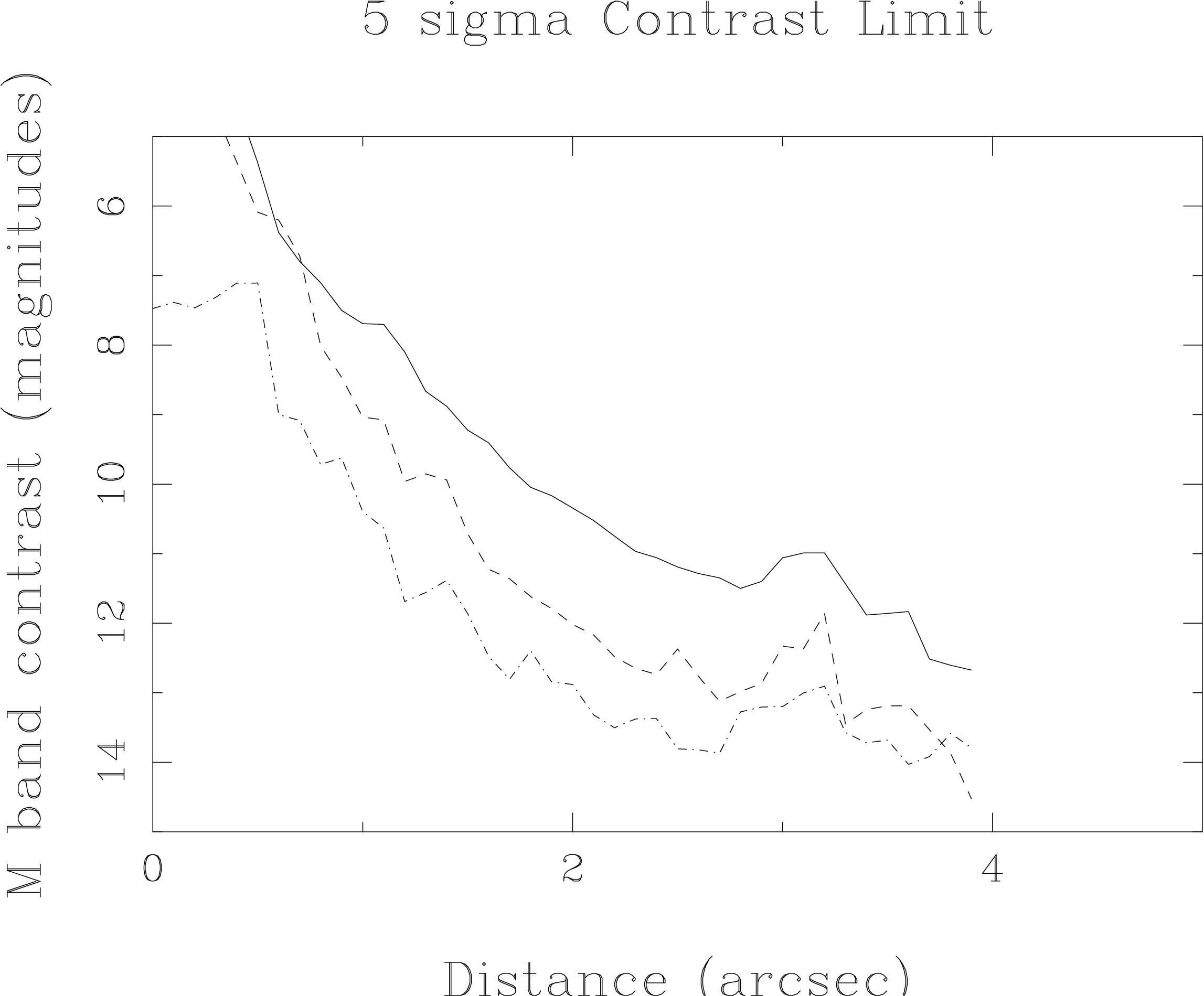}\\
\end{tabular}
\end{center}
\caption[results] 
{ \label{fig:results} \textbf{Left:} Nod-subtracted, stacked, and cropped 3975 second L' image of GL 564 image (shown in logarithmic scale). The first few Airy rings of the primary (G2V star) are visible along with the close-separation twin L-type brown dwarf companions to the lower left. The core is saturated in the central star. The first diffraction ring is visible around the brown dwarf binary, and this binary is visible after a 1.5 s integration. The individual brown dwarfs are not resolved here and are known be 8 H-band magnitudes fainter than the central star and are located 2.64" away~\cite{2002ApJ...567L.133P}. \textbf{Right:} M-band contrast limit versus separation for Vega observation~\cite{HinzApJ}. The approximate contrast limit in M band magnitudes are plotted for Clio on the MMT for a PSF image (solid line), for an unsharp masked image (dashed line) and for a PSF-subtracted image (dash-dot line).}
\end{figure}

\section{The Science}
We carried out our first scientific run in April 2006, and have planned observations in Summer 2006. Scientific projects range from planet searches around most nearby A through M stars and white dwarfs to T-dwarf photometry. Looking for planets around white dwarfs is especially lucrative since these systems offer the best contrast. We are carrying out a massive survey of most nearby moderate age systems. During first light we observed Vega and placed the lowest mass limits for an expected companion~\cite{HinzApJ}. We show the contrast ratios we reached in Figure 4 from our integration. As of now, we have observed 7 stars and have so far not detected any planetary companions. In one case, HD 133002, we found a previously unknown stellar companion. Follow-up imaging is required to establish physical association.  We have successfully tested a PSF shaping phase plate to test out new ways of stellar light suppression (Codona et al. 2006, this conference). Initial on-sky tests show we are able to suppress the parts of the stellar halo by 3 orders of magnitudes through this technique. This will allow close-in imaging of planetary systems.

\section{Conclusion}

The instrument is in a state where it can be used for routine observations and is able to set very interesting new limits in planet detection. We are finally able to do a systematic survey of a wide variety of nearby systems and place mass constraints down to 5 M$_{Jup}$ even for moderate age systems. In the future, we hope to extend the capabilities of our instrument through the addition of L' and M-band PSF shaping phase plates and a grism for crude spectroscopy of our objects. We plan to investigate other coronographic techniques to improve 3-5 micron contrasts.

\acknowledgements
We thank Mitch Nash, Elliott Solheid, and Brian Duffy for assisting with the hardware end of things. We thank Bill Hoffmann for his discussions about Clio electronics. We are grateful for the telescope operators who made our telescope experience an enjoyable one. We would also like to thank the MMTAO team for ensuring the AO system performed spectacularly. 

\bibliography{spie2006}
\bibliographystyle{spiebib}

\end{document}